\documentclass[]{spie}  

\usepackage{amsmath,amsfonts,amssymb}
\usepackage{graphicx}
\usepackage{subcaption}
\usepackage[colorlinks=true, allcolors=blue]{hyperref}

\title{Observers’ Data Access Portal: realtime streaming for astronomical data}

\author[a]{T. Coda}
\author[b]{T. Oluyide}
\author[b]{M. S. Lynn}
\author[a]{J. A. Mader}
\author[b]{G. B. Berriman}
\author[a]{M. Brodheim}
\author[b]{C. Gelino}
\author[b]{J. Good}

\affil[a]{W.M. Keck Observatory, 65-1120 Mamalahoa Hwy, Kamuela, HI, USA 96743}
\affil[b]{Caltech/IPAC-NExScI, 1200 E California Blvd, Pasadena, CA, USA 91125}

\authorinfo{Send correspondence to tcoda@keck.hawaii.edu}

\begin{document} 
\maketitle

\begin{abstract}
 The W. M. Keck Observatory Archive (KOA) has released the Observers’ Data Access Portal (ODAP), a web-application that delivers astronomical data from the W. M. Keck Observatory to the scheduled program’s Principal Investigator and their collaborators anywhere in the world in near real-time. Data files and their associated metadata are streamed to a user’s desktop machine  moments after they are written to disk and archived in KOA. The ODAP User Interface is built in React and uses the WebSocket protocol to stream data between KOA and the user. This document describes the design of the tool, challenges encountered, shows how ODAP is integrated into the Keck observing model, and provides an analysis of usage metrics.
\end{abstract}


\section{INTRODUCTION}
\label{sec:intro}  

\begin{figure} [ht]
\begin{center}
\begin{tabular}{c} 
\includegraphics[height=8cm]{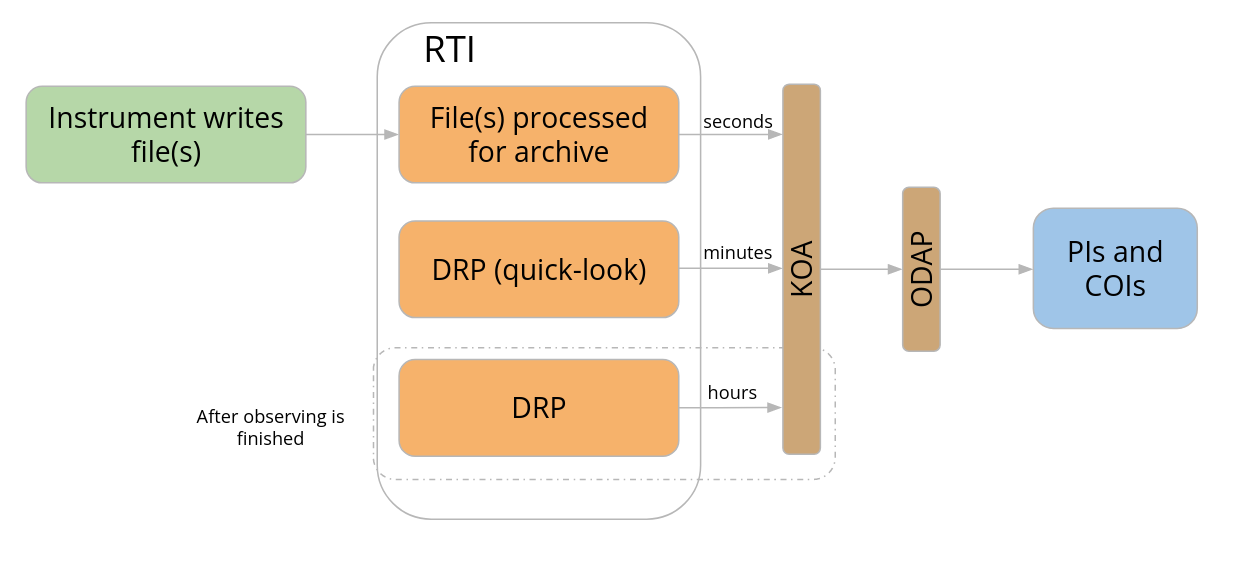}
\end{tabular}
\end{center}
   \caption[WMKO observing model] 
   { \label{fig:wmko_observing_model} 
 Flow chart of the  Keck Observing model. Starting from left to right, the instrument writes files to a directory, which KOA/RTI processes as a raw data product. Files are then passed onto KOA and the DRPs. During an observing night, ODAP interfaces between KOA, and the PI and their observing team.
}
\end{figure}

As part of the Data Services Initiative project, The W. M. Keck Observatory (WMKO) has recently overhauled their observing model such that data reduction and archiving are tightly coupled with nightly observing\cite{brodheim2022}. Figure \ref{fig:wmko_observing_model} depicts the new flow of data from creation by the instrument software, to the archiving of both raw and reduced data products. Additionally, automated Data Reduction Pipelines (DRPs) supply data product files to the archive. Implementing the Real-Time Ingestion (RTI) process improved the data transfer process over to the Keck Observatory Archive (KOA) \cite{berriman2022realtime}, with files being transferred with seconds of creation. Prior to RTI, principal investigators (PIs) had to wait until the following afternoon to get their data from the KOA. With these improvements, PIs and their collaborators now have access to their data immediately upon ingestion at KOA, allowing them to quickly analyze the data and make changes to their observing strategy during the night.

Access to these data used to be provided in two ways: the Observer would download the raw data in bulk from storage at WMKO or the KOA web-based user interface was be used to search for both raw and reduced data. Both of these methods were not adequate for use while observing as the former only allowed access to raw data and users could potentially access other group's data, and the latter required a separate search request as new data were archived.

\begin{figure} [ht]
\begin{center}
\begin{tabular}{c} 
\includegraphics[height=9cm]{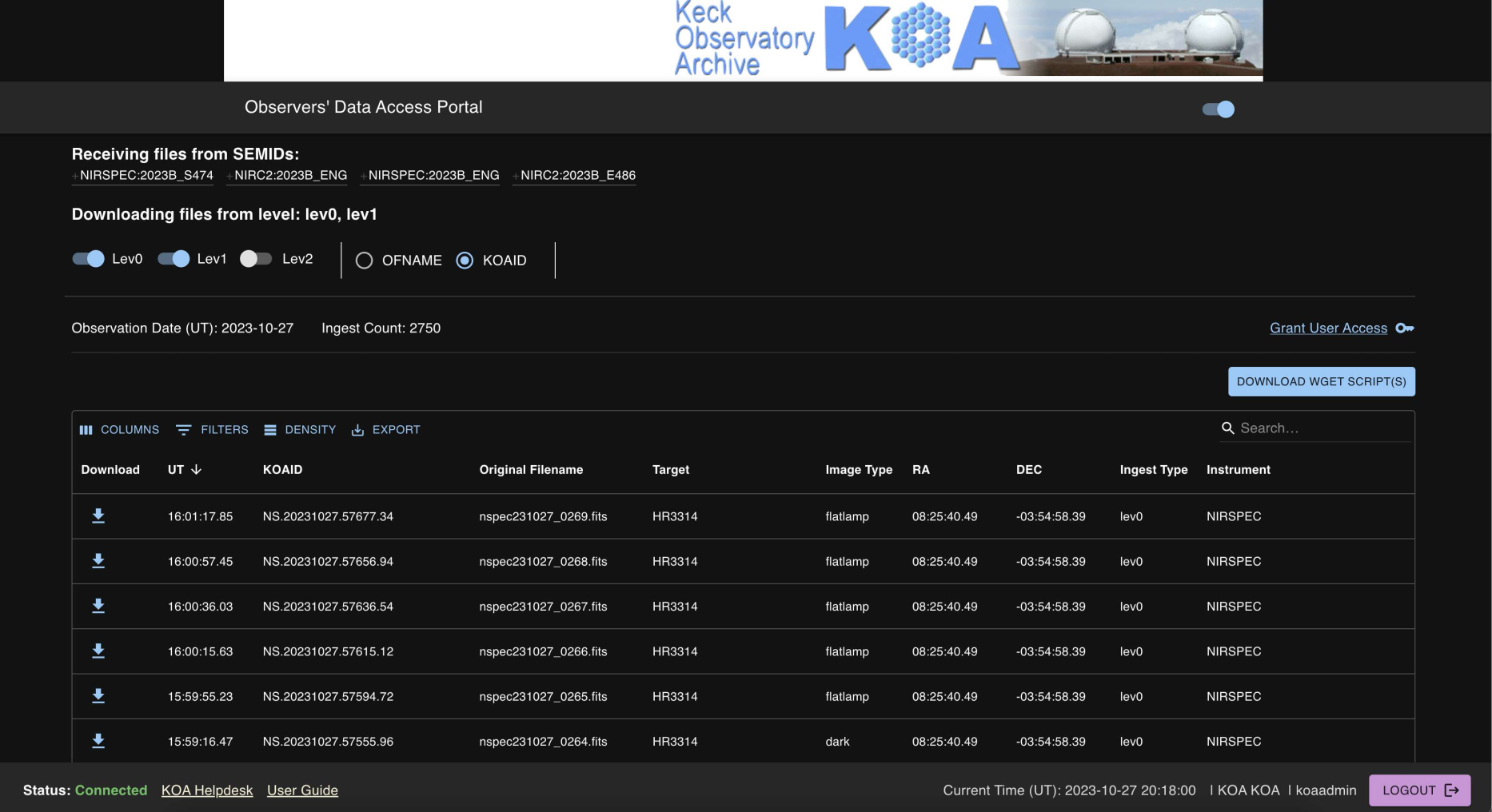}
\end{tabular}
\end{center}
   \caption[odap screenshot] 
   { \label{fig:odap_screenshot} 
Screenshot of the Observers’ Data Access Portal
}
\end{figure}

To streamline this process and to provide access to all data products seamlessly while observations are underway at night, WMKO and IPAC-NExScI have developed and released the Observers' Data Access Portal (ODAP).\cite{oluyide2024observers} ODAP seen in Figure \ref{fig:odap_screenshot} builds upon the work done by RTI \cite{berriman2022realtime}, improving the data delivery aspect of the WMKO observing process, providing observers near-realtime access to raw and reduced data products from KOA.

\section{Overview of the Observers' Data Access Portal}

The ODAP user interface (see Figure \ref{fig:odap_screenshot}) is a web-application based on React, a JavaScript framework written to create complex user interfaces.  ODAP is written in TypeScript, a super-set of the JavaScript language, which extends JavaScript to include typing.  While using TypeScript often results in additional development time, its use allows for making the code base more human readable, and improving debugging/maintenance efforts, often saving time and resources in the long run.  The web-application uses WebSockets, a standard channel of asynchronous communication and data transfer between a client and server, to receive data from KOA.  WebSockets are implemented to update the metadata table and download files as they are ingested by RTI.

Upon logging in, ODAP verifies that the user is associated with a scheduled observing program and ensures that they will only have access to the data for which they have permission. As new data are ingested by RTI, the file’s metadata is broadcast to the subscribed user’s instance of ODAP and displayed on an interactive table specific to the night of observing. If enabled, incoming files are automatically streamed to the user’s computer, allowing them to inspect their data on their desktop moments after the data were written to disk at KOA.

\section{Architecture}

\begin{figure} [ht]
\begin{center}
\begin{tabular}{c} 
\includegraphics[height=7cm]{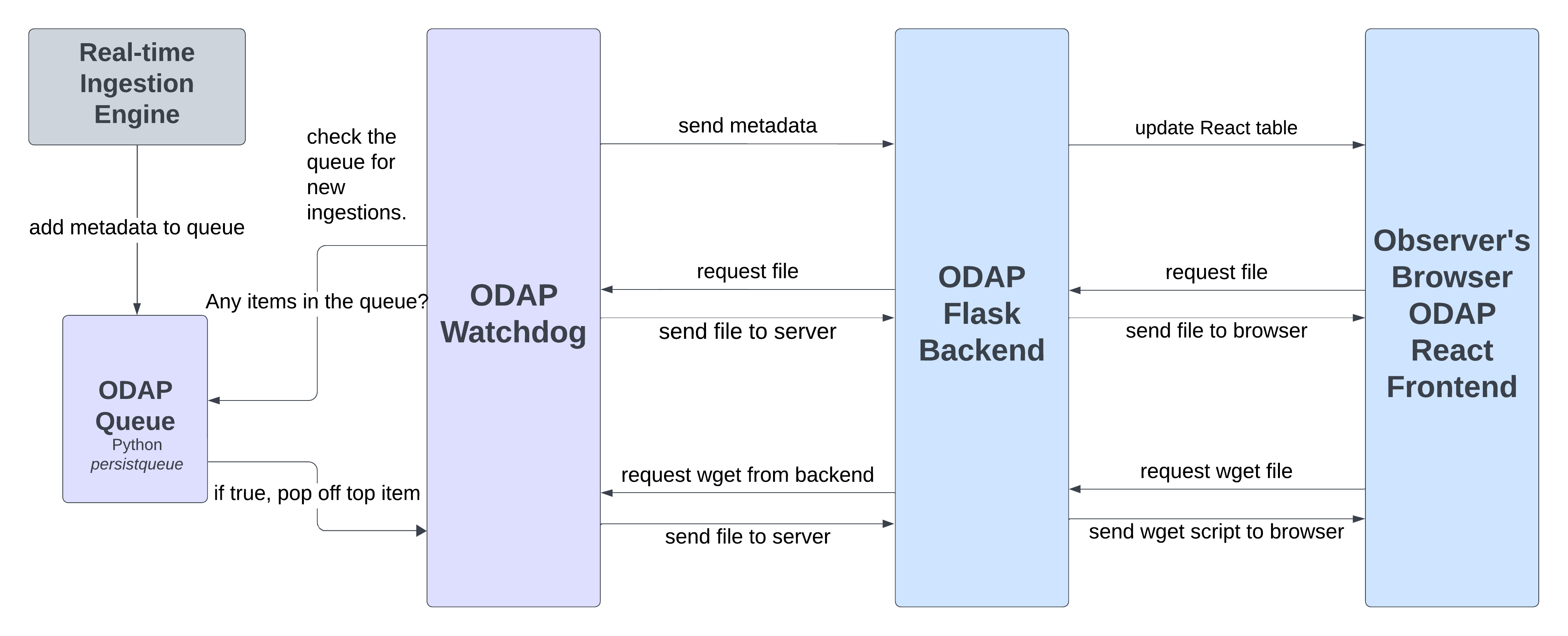}
\end{tabular}
\end{center}
   \caption[odap flowchart] 
   { \label{fig:odap_flowchart} 
 Diagram showing the flow of data from RTI to the observers’ browser. 
}
\end{figure}

 ODAP consists of three components, a front-end client, a back-end server, and a watchdog client. Together, these components pass metadata that record file ingestion events from RTI to the Observer. The RTI interfaces with the watchdog by writing ingestion event data to a disk-stored, thread-safe queue. The watchdog client checks for events in the queue. If any are found, it removes and forwards the events to the Python Flask back-end server, which then proxies them from KOA to the front-end client on the observers' browser. A diagram describing the architecture is given in Figure \ref{fig:odap_flowchart}.

Files can then be downloaded, either automatically or by request. If automatic download is enabled, the GUI requests a file. That request is proxied to the Watchdog, who has full access to the files. The response is returned to the user and the requested file is downloaded by their browser.

Depending on file/size and quantity, downloading the whole night's payload in one request can overwhelm ODAP. To prevent a decrease in performance, ODAP downloads files one-by-one while the bulk download options is handled by directly calling KOA's API via WGET scrips

\subsection{Watchdog}
\label{sec:watchdog}

 The watchdog interfaces between the RTI and the back-end server. The watchdog process monitors newly ingested files on KOA’s data disks. The RTI system ingests a file and records the event in a persistent queue, which is stored in a SQLLite3 database on disk. The watchdog checks the queue for events, and when one is found, sends a message the back-end server with the file’s metadata, which in turn is forwarded to the front-end client(s). The watchdog also responds to requests coming from the back-end server, such as fetching files or gathering metadata for all previously ingested files. Once the entry is processed, it is removed from the queue.

\subsection{Back-end Server}
Clients (watchdog and front-end client) use the back-end to pass messages to each other, allowing one client to indirectly invoke the other(s) to perform an action. For example, the front-end client requests all ingestion event metadata from the watchdog, a message is passed to the server, and then a separate message is passed to the watchdog. The watchdog then carries out the request, sending back the metadata to the server, which is then passed onto the client who made the request.
The decision to have the watchdog and back-end run on separate processes compartmentalizes access. The watchdog having access to all of KOA, shares only what is requested by the back-end. The back-end in turn verifies the user is authorized to request data.

\subsection{Front-end Client}
 The front-end GUI is the client running on observers' web browser that they use to receive metadata/files. The client starts the sign-in handshake with the back-end server using the observer’s KOA account. Upon a successful login, an API token is provided to connect the front-end client to the back-end server \cite{oluyide2024observers}. The user is then authorized to receive program data they are associated with. Users are provided a list showing which semester IDs they are assigned in the GUI, as well as an interactive table showing ingested files, shown in Figure \ref{fig:odap_screenshot}. The user may sort, search, and filter this table. Additional settings provide control over which files (none, raw, and/or reduced) are automatically downloaded.

\begin{figure*}
    \centering
    \begin{subfigure}{1\textwidth}
        \centering
        \hspace{1.25in}
        \includegraphics[height=5cm]{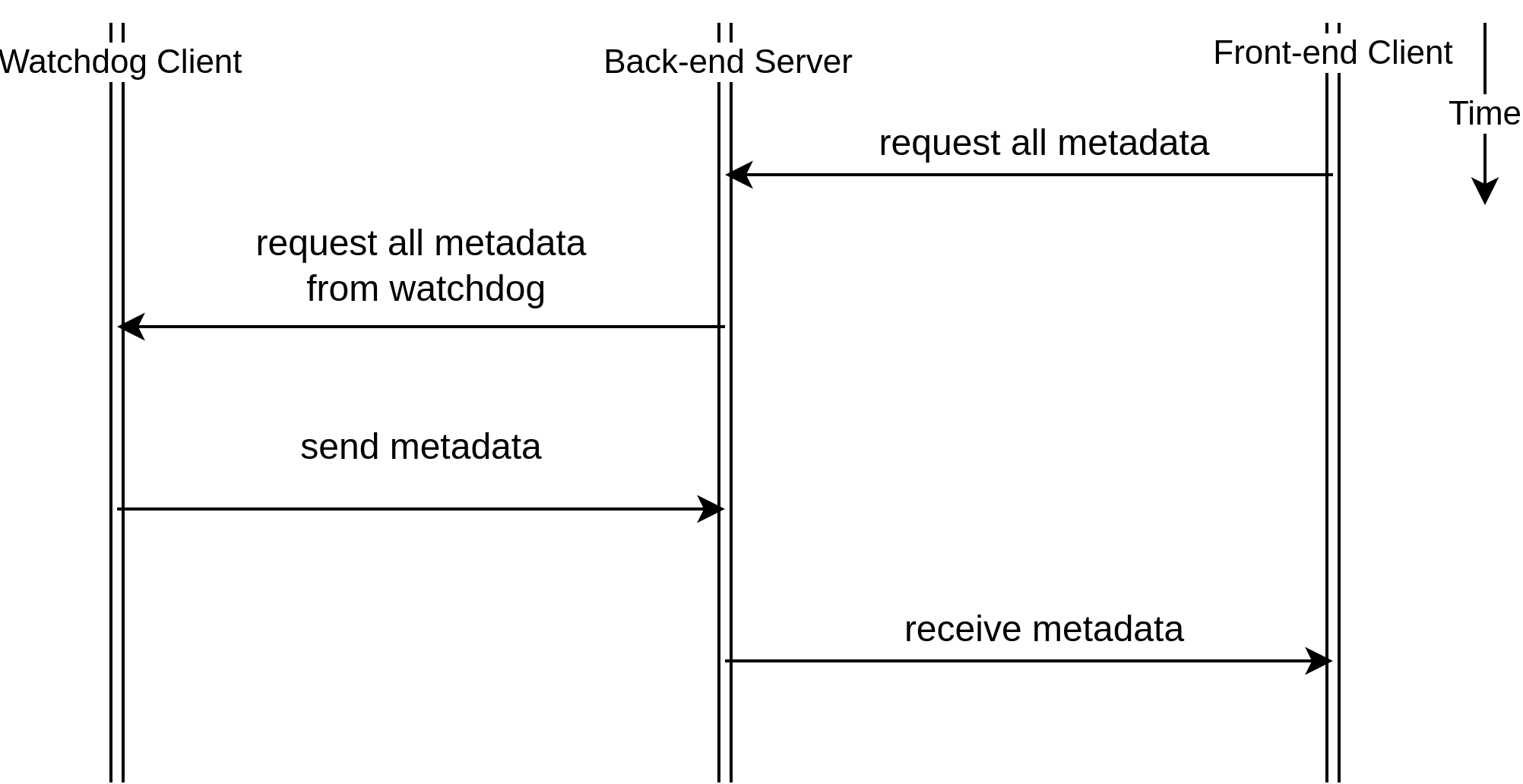}
        \caption[metadata event] 
   { \label{fig:mdfc} 
 Sequence diagram of metadata request. This request is carried out each time the interface is opened.}
    \end{subfigure}
    \begin{subfigure}{1\textwidth}
        \centering
        \includegraphics[height=5cm]{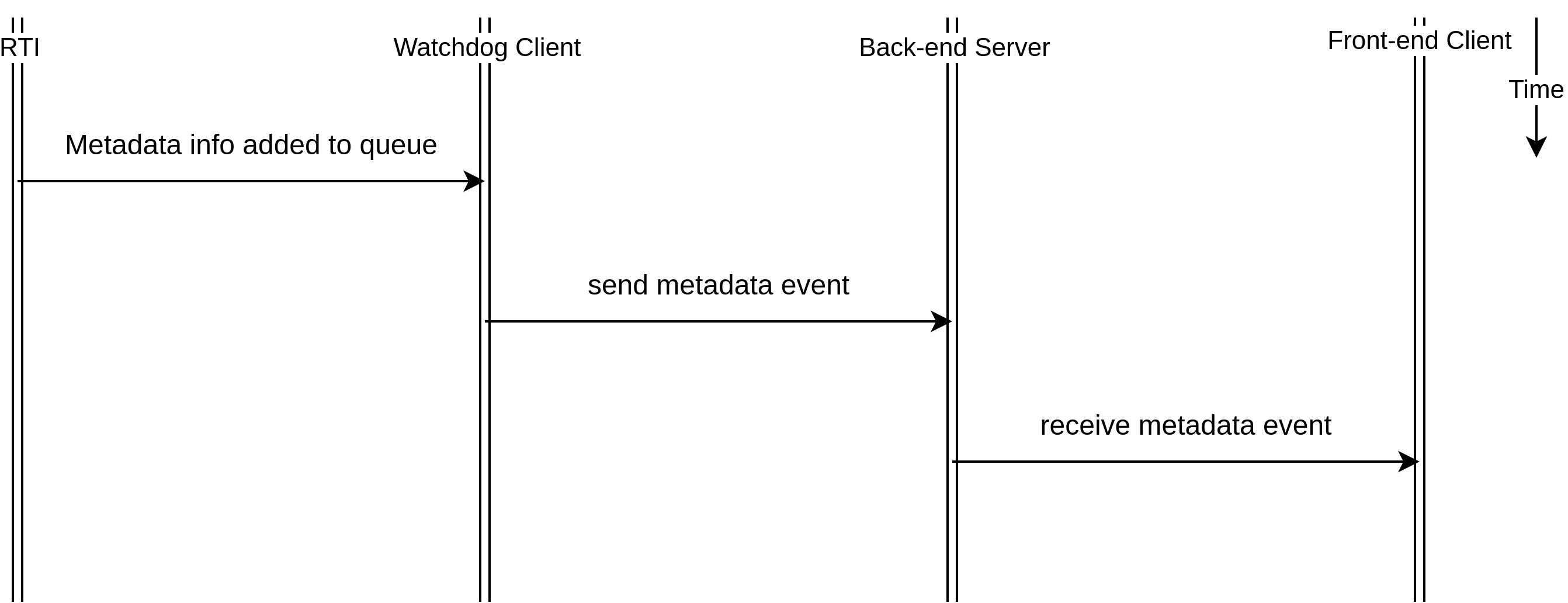}
        \caption[broadcast event] 
   { \label{fig:bre} 
Sequence diagram of an RTI ingested file's metadata to the ODAP GUI.
}
    \end{subfigure}
    \begin{subfigure}{1\textwidth}
        \centering
        \hspace{1.25in}
        \includegraphics[height=5cm]{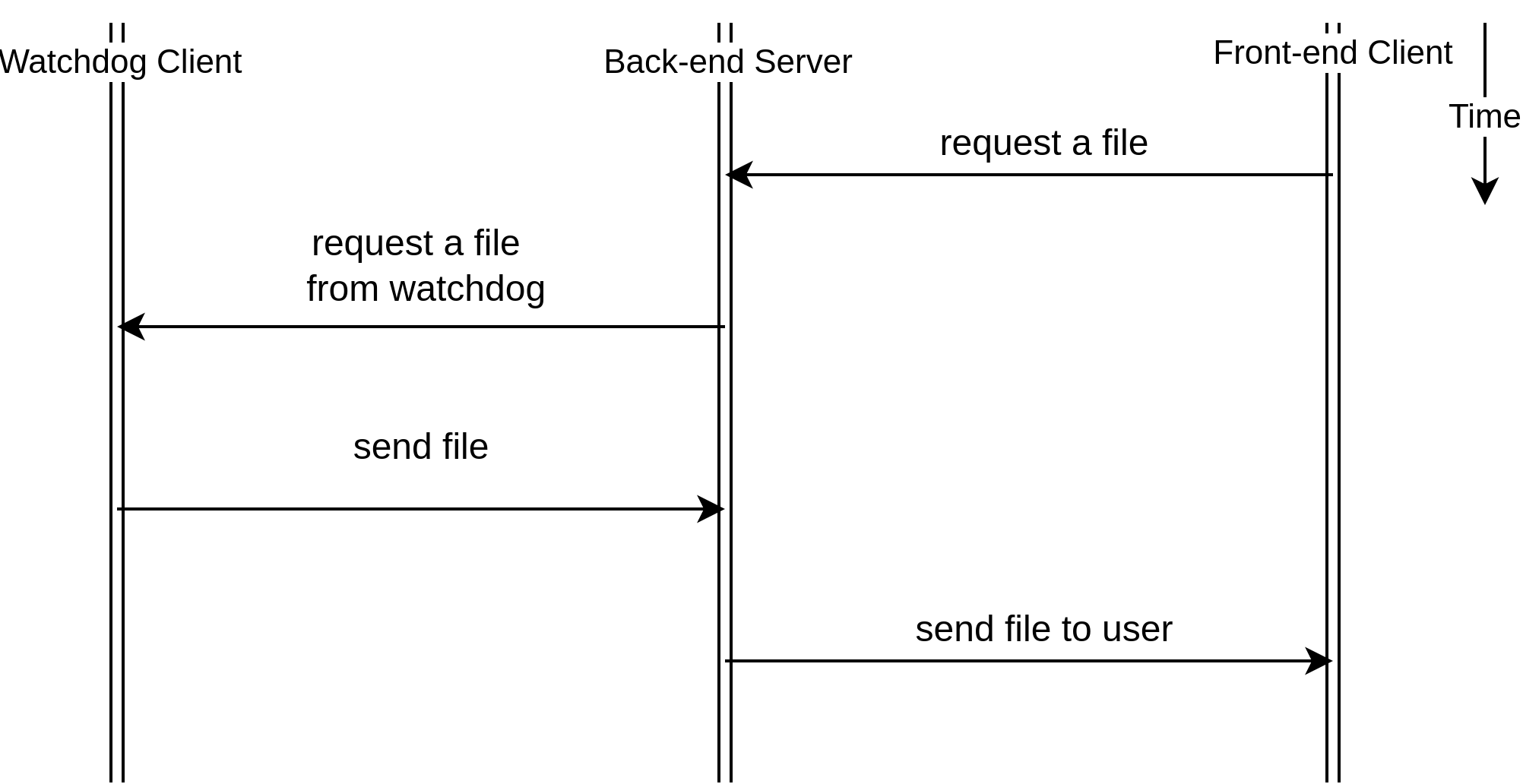}
        \caption[metadata event] 
   { \label{fig:ffc} 
 Sequence diagram of file request. This request is carried out each time the user requests a file.}
    \end{subfigure}
    \caption{
    \label{fig:fc}
    ODAP Sequence diagram between clients and server. Events are represented as horizontal lines. Software components are represented as vertical lines. Time moves downwards.}
\end{figure*}

Metadata is retrieved in the manner shown in Figure \ref{fig:mdfc}. The front-end client sends a request to the back-end server asking for all of the programs' metadata already ingested upon loading the page. The back-end server response contains all relevant metadata from the watchdog. An identical procedure is made for sending files shown in Figure \ref{fig:ffc}.

The RTI initiates the the broadcast of newly ingested files in the process covered in the \ref{sec:watchdog} subsection. File ingestion event metadata is sent to the front-end client as shown in Figure \ref{fig:bre}. Upon receipt of this message, the front-end client updates its table with a new row. If downloading is enabled, receipt of a new row generates a follow-up request to download the ingested file using the process mentioned in Figure \ref{fig:ffc}.

\section{IMPLEMENTATION}

\subsection{Scalability and Performance}
File sizes of the data produced by WMKO instruments are in the range of 10 to 100's of megabytes, with new data formats reaching as high as 500 megabytes. The files are compressed and transported via WebSockets, which has no in-built size limitation. With this use case in mind, it was found that ODAP is able to serves these files. 

\subsection{Data Access}
KOA data access for an individual observing program is only granted for the PI of the program.  An account and data access rights are created before the observing run starts.  The PI must grant data access to other users' KOA accounts.  This request can be made either through the KOA help desk or by using a web form.  The creation of non-PI accounts must be requested through the help desk before the help desk closes at the end of IPAC-NExScI's business hours.



\subsection{Use Metrics}
Google Metrics is used to gather user data. ODAP was released on August 1, 2023 and as of May 8th  there have been 550 users who have accessed ODAP. 505 of these users are located in the USA. The time the user is actively using the GUI, defined as the time the browser's tab serving ODAP is in focus, is averaged at 3 minutes 41 seconds. ODAP is still able to stream files while it is not in focus. Adding metrics to record the number of files being streamed is an ongoing improvement.

\subsection{Challenges Encountered and Lessons Learned.}

At the start of the project, developers took the lead on writing requirements, and would be presented to the stakeholders, consisting of PIs, and Staff Astronomers, and KOA staff. As a result, requirements were focused primarily on technical implementation and not typical astronomer workflow. In hindsight, having the developers write the requirements led to too much focus on technical needs and less on the needs of the stakeholders (WMKO observers, staff astronomers and KOA operations). Originating requirements from the stakeholder would have been easier to refine into technical requirements for the development team. 

One oversight was found in the requirement “...[ODAP] shall provide a link to the KOA Help Desk, which will be staffed during California and Hawaii business hours” Many observing team members were denied access to their data upon logging in after business hours at the start of observing. At that point it was too late for KOA Help Desk to fix the problem. The issue was addressed by sending the PI emails reminding them to authorize their observing teams' accounts. An additional solution has the PI fill out a web form that performs the observing team authorization requests automatically, without the need for the KOA help desk, but as of now is not in operation due to concerns over user account security.

Developer-driven requirements increases the risk of interpretations to vary between the stakeholders. The requirement that ODAP users “...shall use their established KOA or Keck Observatory credentials to access this [ODAP].” is vague, and should have been refined once it is clear who should host ODAP: WMKO or KOA. The decision to access using a KOA account to use operational software brought unintended consequences when creating and managing two accounts. Furthermore the requirements did not address the policy for KOA to create accounts only for PIs. The Observing team needs to request the KOA Help Desk during business hours for account creation. Too often, an observer who has access to WMKO operational software could not view the data they were taking because their KOA account either did not exist or was not granted access beforehand.

Difficulties in addressing access challenges quickly were exacerbated due to design decisions made early on in the project. The front-end development was done in React using TypeScript, which some developers were not familiar with. Having joined much later, the decision to use React/Typescript did not consider the practicality of developers having the experience to develop with the required skills in mind. Thus, instead of developers addressing problems in implementation, they were thrust into an unfamiliar language using an unfamiliar toolkit. React/Typescript training beforehand would have given the new developers the skill set needed to address access challenges previously mentioned.

Had the requirements initially originated from the stakeholders as opposed to the developers, they could have perhaps foreseen the challenges of duel accounts that were created and authorized outside of observing hours. Nevertheless, the release of ODAP without an comprehensive authorization process in place marred many users' first impression of the tool when they were denied access. In response, the KOA team designed an automated account creation/authorization utility; however, account security concerns has stalled a solution satisfactory to WMKO Observers. 

\subsection{Future Enhancements}
ODAP will continue to be updated as observers provide feedback on its use.  For the user experience, future enhancements include:
\begin{itemize}
\item{Access to data for non-classically scheduled observation like Target of Opportunity and twilight observations.}
\item{Adding a React FITS viewer with features such as scaling, zooming, panning, and rotation.}
\item{Adding a standard image viewer to display quality assurance plots output by the DRPs.}
\item{Automate data access permission to ensure usage by all program team members.}
\end{itemize}

Additional features will also be added to help record use metrics:
\begin{itemize}
  \item{Include ways to record the number of downloads that are initiated from ODAP.} 
  \item{Improve and extend the logs on the back-end server.} 
\end{itemize}

\section{CONCLUSION}
ODAP’s position in the WMKO observing model has seen success in delivering observers their data during observation, when they have been properly authorized. A focus on technical requirements instead of observer workflow led indirectly to authorization challenges that are continuing to be addressed; once resolved, ODAP shall be refined further to suit the needs of WMKO observers.

\acknowledgments 
 
Some of the data presented herein were obtained at Keck Observatory, which is a private 501(c)3 non-profit organization operated as a scientific partnership among the California Institute of Technology, the University of California, and the National Aeronautics and Space Administration. The Observatory was made possible by the generous financial support of the W. M. Keck Foundation. 

The authors wish to recognize and acknowledge the very significant cultural role and reverence that the summit of Maunakea has always had within the Native Hawaiian community. We are most fortunate to have the opportunity to conduct observations from this mountain.

\bibliography{main} 
\bibliographystyle{spiebib} 

\end{document}